# Emotion recognition using a glasses-type wearable device via multi-channel facial responses

Jangho Kwon[1,2], Laehyun Kim[2,*]

[1] Hanyang University; jhkwon@spa.hanyang.ac.kr (J. K), junwchoi@spa.hanyang.ac.kr (J. C)
[2] Korea Institute of Science and Technology; hj910410@kist.re.kr (J. H), dahyekim@kist.re.kr (D. K)
* Correspondence: laehyunk@kist.re.kr; Tel.: +82-2-958-6726

**Abstract** We present a glasses-type wearable device to detect emotions from a human face in an unobtrusive manner. The device is designed to gather multi-channel responses from the user's face naturally and continuously while he/she is wearing it. The multi-channel responses include physiological responses of the facial muscles and organs based on electrodermal activity (EDA) and photoplethysmogram. We conducted experiments to determine the optimal positions of EDA sensors on the wearable device because EDA signal quality is very sensitive to the sensing position. In addition to the physiological data, the device can capture the image region representing local facial expressions around the left eye via a built-in camera. In this study, we developed and validated an algorithm to recognize emotions using multi-channel responses obtained from the device. The results show that the emotion recognition algorithm using only local facial expressions has an accuracy of 78% at classifying emotions. Using multi-channel data, this accuracy was increased by 10.1%. This unobtrusive wearable system with facial multi-channel responses is very useful for monitoring a user's emotions in daily life, which has a huge potential for use in the healthcare industry.

**Keywords:** wearable device; emotion recognition; affective computing; facial response; facial expression;

## 1. Introduction

Emotion recognition is a technology to predict people's emotional states based on user responses such as verbal or facial expressions; this technology can be applied in various fields, such as health care [1], gaming [2], and education [3]. To aid these applications, the technology should recognize emotions in real-time and naturally while the user is experiencing them. Recently, wearable devices have garnered attention for emotion recognition applications [4].

Most existing wearable devices for emotion recognition have relied on bioelectrical signals. The bioelectrical signals are electronic signals that indicate physiological responses, such as a pulse or a sweating response. These signals originate from changes in the autonomous nervous system (ANS), which is a control system that regulates human bodily functions. With regard to the ANS, there is a risk of misunderstanding a user's emotional state if the wearable devices relies on the biosignals alone, because the ANS is affected not only by emotion but also by other factors, including cognitive stress [5][6] or physical activities [7]. For instance, many people in workplaces experience physical activities and cognitive stress, which affect their biosignals; therefore, using only the bio-signal may not be reliable [6]. In this case, it would be desirable to use additional modalities to obtain more reliable emotional information.

Facial expression can be used as an additional modality here as it provides important cues for emotion recognition. This modality is already used in studies on wearable devices [7]. There are two methods for extracting facial expressions via the wearable devices. The first is the sensor-based method, which measures the movements of facial muscles to reflect emotions [8][9]. This approach might cause discomfort owing to contact with the skin on the facial muscles. The other method is the camera-based method, which captures facial expressions using a camera [10]. This method has advantages over the sensor-based method because the camera is not attached to the skin. Nevertheless, camera-based methods have been not used frequently in wearable devices because the modules were cumbersome and heavy to wear [11]. However, owing to advancements in technology, the sizes of camera modules have become small enough to wear comfortably.

Recently, camera modules have been used in commercial wearable devices. In particular, they have been primarily used in head-mounted wearable devices to capture the user’s perspective, and there have also been emotion recognition studies using the captured pictures [12],[13]. However, these studies used pictures to recognize the emotions of other persons and not those of the users. We assume that it is beneficial to use camera modules for monitoring the user’s own emotions using these glasses-type wearable devices, and this can be achieved easily by turning the direction of the mounted camera.

There are some issues associated with turning the camera module. Owing to the short distance between the face and the camera, the wearable device can capture only a portion of the facial expression. Additionally, to avoid blocking the user’s view, their expression should be captured from the side. Therefore, the acquired data would be not typical for emotion recognition studies, so it is important to check the feasibility of whether partial and side-facing expressions can be used to monitor emotional state.

In summary, to check the feasibility of our idea, we propose the following two hypotheses.

The local, side facial expression can be used to monitor a user’s emotional state.

Combining local facial expressions and physiological responses yields a better outcome than using physiological responses alone.

To test this hypothesis, we developed a glasses-type wearable device that uses multi-channel facial responses, which include local facial expressions and physiological responses. Using the developed device, we conducted an induced emotion recognition experiment with emotional video clips. The experimental results demonstrated that we were able to predict the four induced emotions with an accuracy of 78% using only hand-crafted features and data from the developed wearable device.

Our study contributes to the use of wearable devices for emotion recognition studies. To the best of our knowledge, a glasses-type wearable device that uses multi-channel facial responses has not been presented by any other research group. The proposed method provides more accurate emotion recognition for complex tasks, such as watching videos. This work may contribute to achieving actual emotion recognition in practical situations.

## 2. Methods

### *2.1. Hardware Implementation*

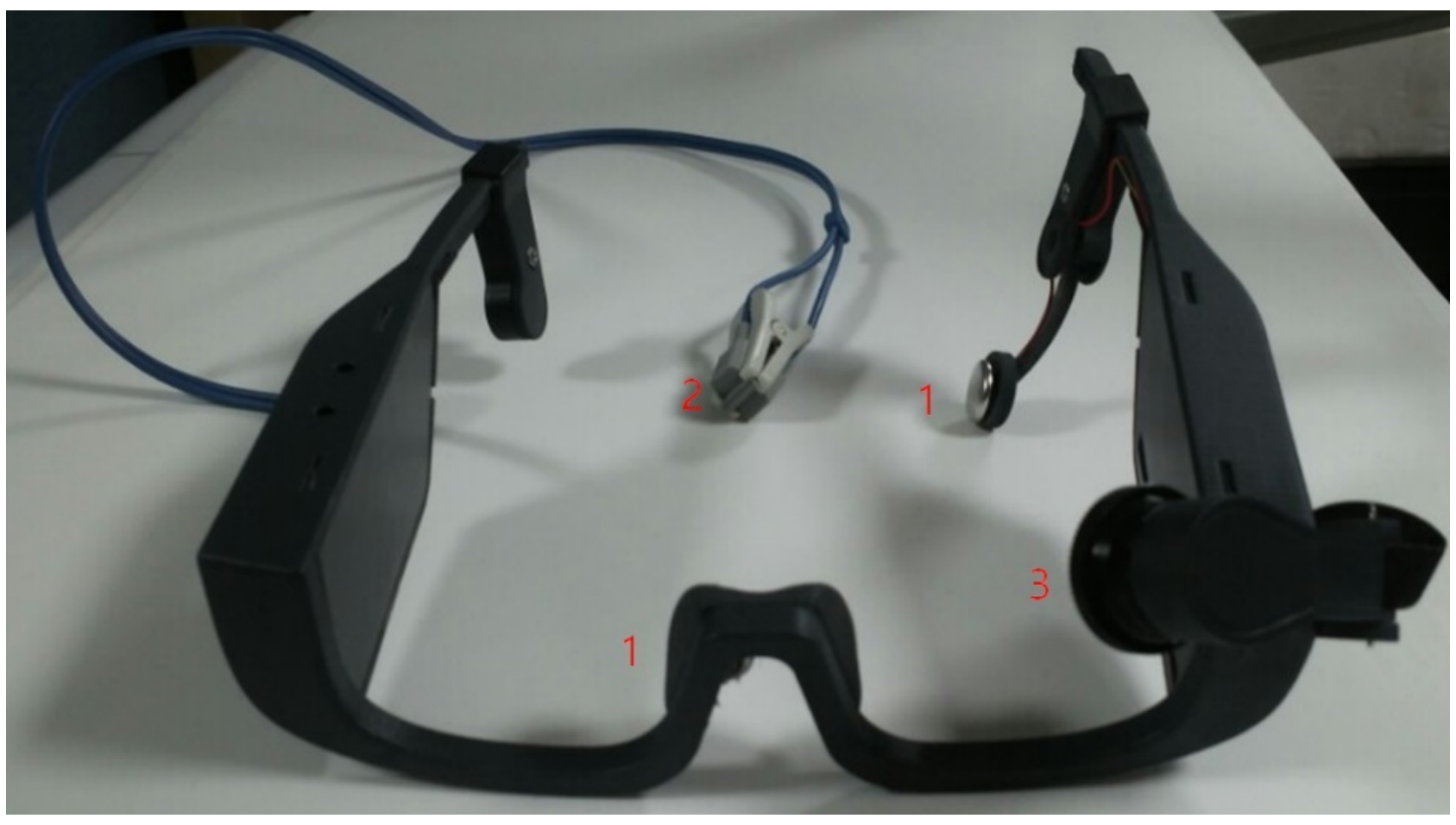

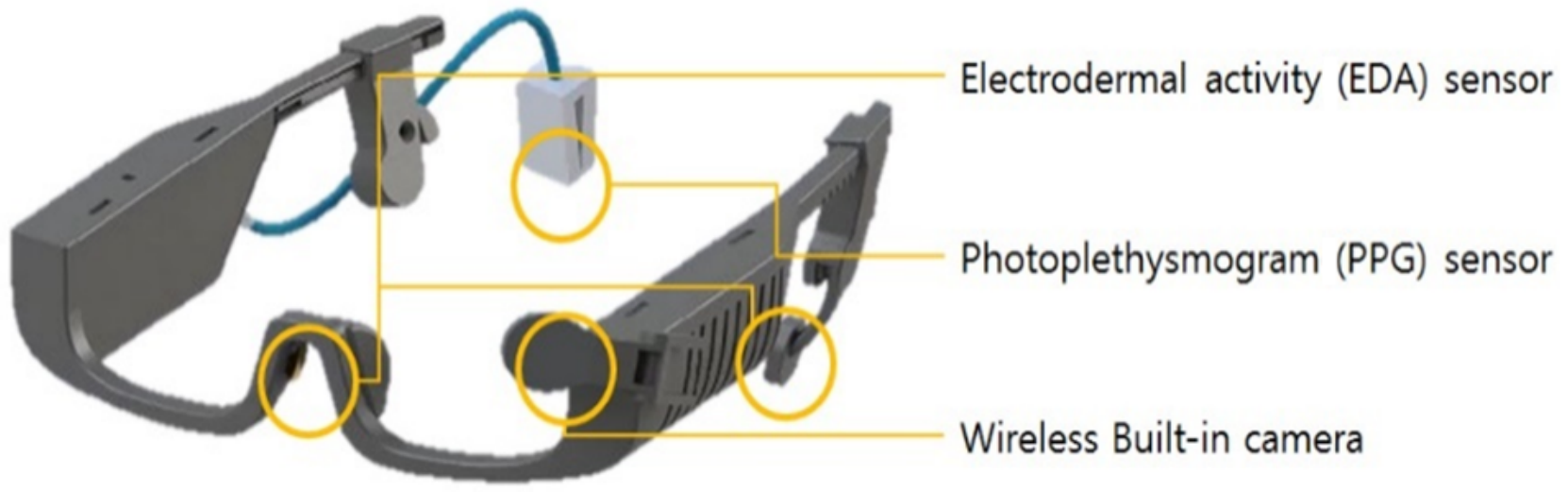

**Figure 1.** Proposed hardware. (1) EDA sensors that measure the skin conductance from the user's nose and mastoid. (2) Ear-clip-type PPG sensor that measures the pulse from the user's earlobe. (3) Built-in camera that measures the user's facial expressions around the eyes.

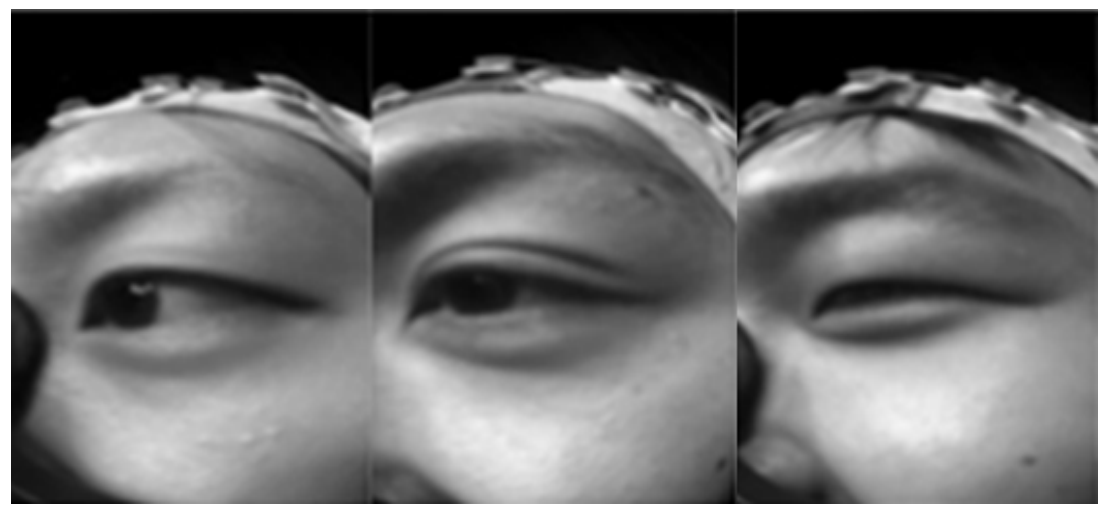

**Figure 2.** Samples of local facial expressions obtained using built-in camera in the wearable device

The multi-channel wearable device for emotion recognition was designed to extract facial expressions and physiological responses. To acquire measurements easily, the device was designed in the form of glasses-type wearable, similar to Google Glass [14] or the prototype sunglasses for emotion recognition presented in CES 2017 [15]. The camera module was attached to the left side of the device to capture the local facial expressions around left eye (from the eyebrows to the cheeks). To measure the physiological responses, electrodermal activity (EDA) sensors were attached to the surface of the skin in contact with the nose and mastoid. Plethysmography sensors were also attached to the earlobes, which have been frequently used in previous studies [16][17][17][17][18]. The concept of the required hardware was presented in our preliminary study [19].

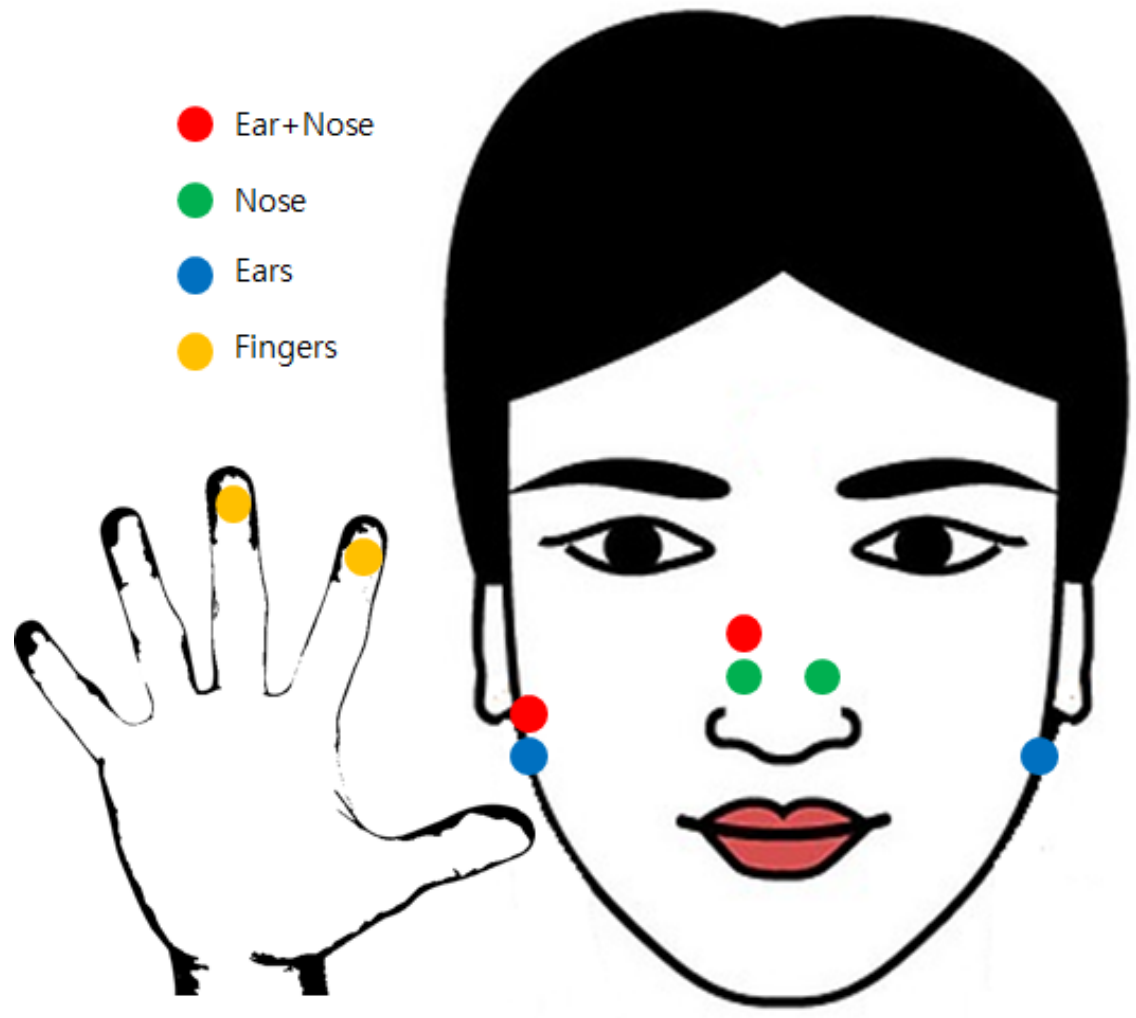

**Figure 6.** Candidate positions for EDA sensor placement.

The position of the EDA sensors was more carefully determined than the other sensors. Sweat glands, which are the sources of the EDA signal, are distributed with different densities on the body [20]. Therefore, it is necessary to find the best location to achieve contact with the EDA sensors. Three candidate positions were compared. The first position is on both sides of the nose. This position can be contacted by the nose-supporting part of the glasses. The second is on both sides of the mastoid, which is just behind the ears. This position can be contacted by the earpiece of the glasses. The last position is a combination of the ear and mastoid. The EDA signal measurement experiment was analyzed to compare these positions. The experiments were performed with 10 subjects. The subjects sat on a chair 90 cm away from a 19-inch LCD monitor. EDA sensors were placed in the three candidate positions and on the left index and middle fingers, and the subjects were requested to watch movie clips for approximately 90 minutes. The EDA signals were measured simultaneously from all the positions while the subject watched the movie clips. All the procedures in the experiment were certified by the Institutional Review Board of KIST-2016.

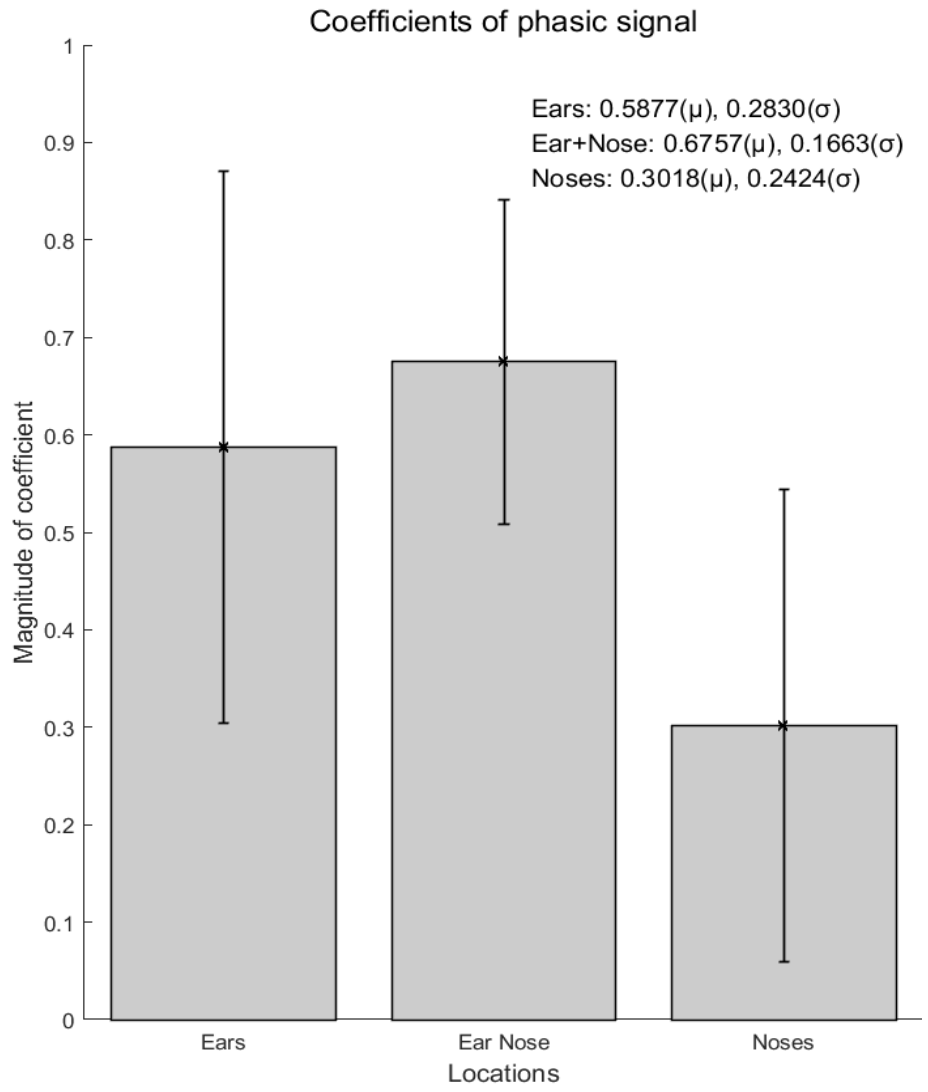

**Figure 7.** Correlation coefficients between fingers and candidate facial sensor placements.

The results of the EDA measurement experiments show that the highest correlation with fingers exists when the sensors are placed near the nose and mastoid. The average correlation from the mastoid and nose is 0.6757, which is higher than those in other locations, and the standard deviation between subjects is 0.1663, which is lower than those in other locations.

## *2.2. Induced Emotion Experiment*

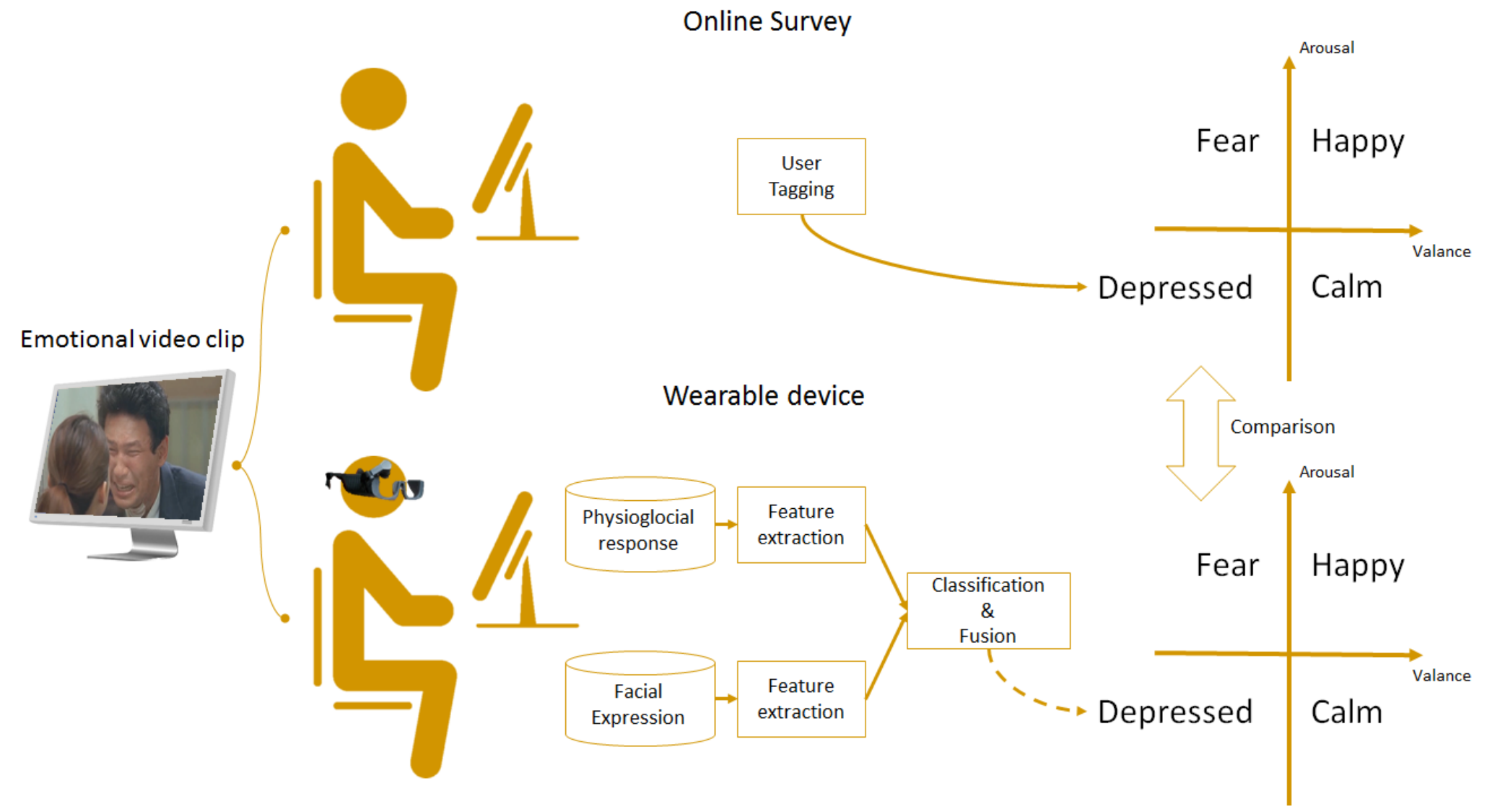

**Figure 3.** Overall flow diagram of the induced emotion recognition experiment.

- User Tagging

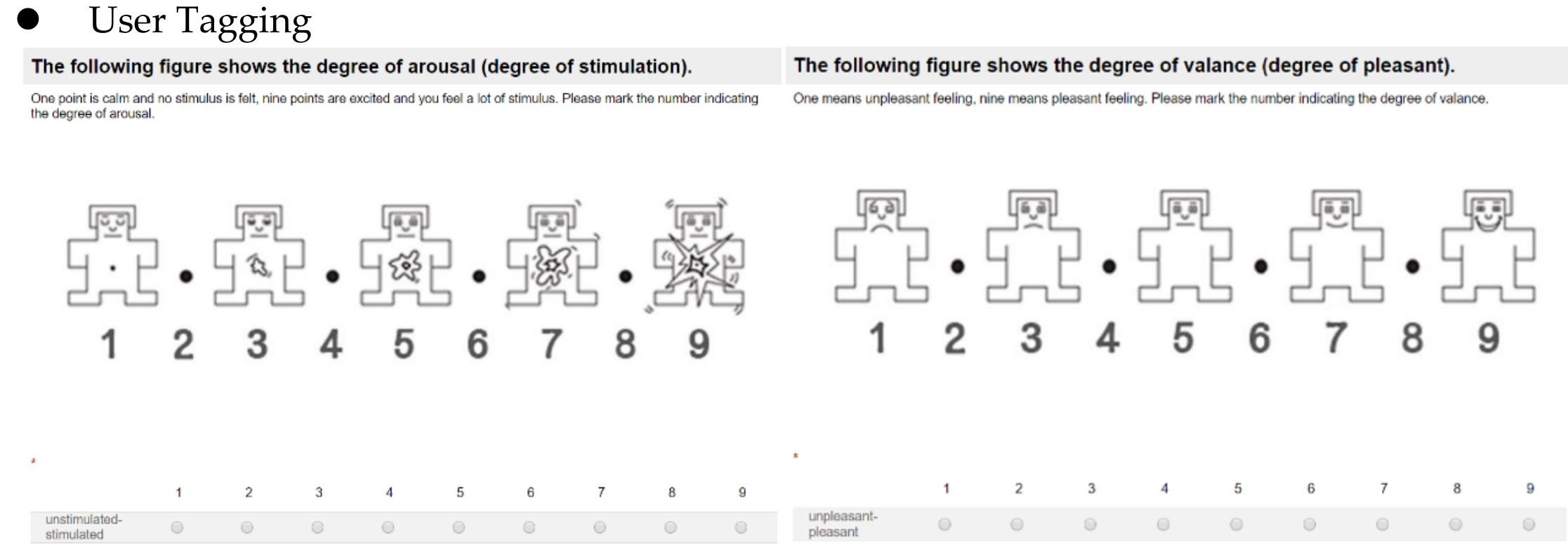

**Figure 4.** Questionnaire form to evaluate arousal and valence resulting from emotional videos in online survey.

We prepared a video-clip-based stimulus to induce emotion in the subjects. We intended to induce the four-dimensional emotional state [21]. The target emotional states include high arousal with high valence (HAHV), high arousal with low valence (HALV), low arousal with low valence (LALV), and low arousal with high valence (LAHV), which correspond to each quadrant of the arousal–valence plane [22].

Two-minute movie clips were used as stimuli to induce emotions. All the clips were extracted from Korean movies, because the subjects are familiar with the language of the clips. First, the clips were carefully selected by five research managers. Twelve movie clips were selected for each category of emotions. Next, we recruited sixty subjects for the online survey, and the degree of the emotion of the clips was ranked by each subject to determine if the stimulus induced strong emotions. During the survey, each clip was watched and given arousal and valence scores between 1 and 9 by each subject. The clips were sorted according to the distance from the origin, and two clips were excluded, which were close to the origin. As a result, a total of 32 video clips were selected.

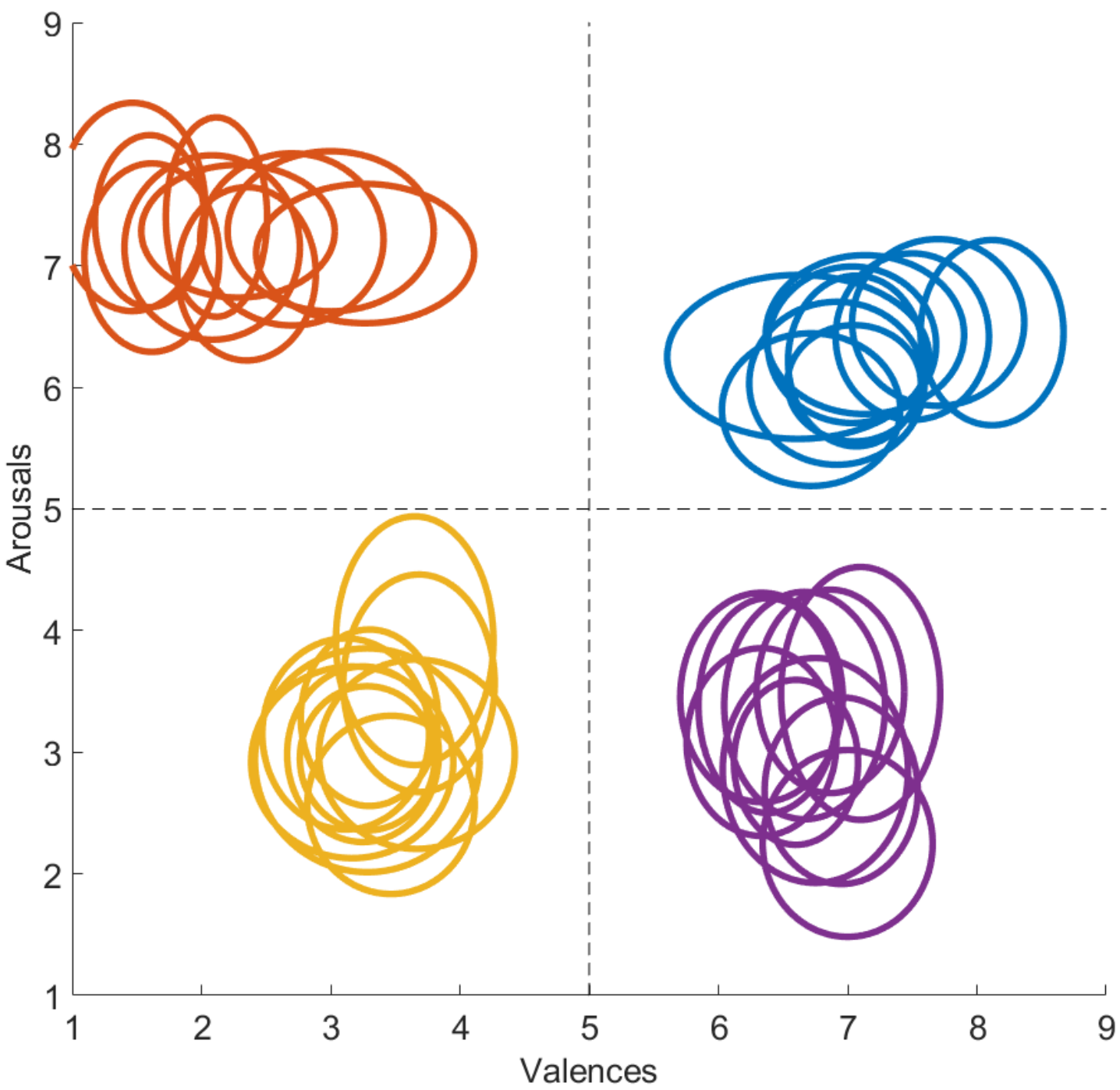

**Figure 5.** Distribution of SAM scores on arousal–valence plane based on survey results.

Figure 5 shows the SAM scores obtained from the online survey. Each circle represents the video clip. The position of the circle indicates the average SAM score of the clip, and the width and height of the circle represent a standard deviation of the valence and arousal scores of the clip, respectively. The distance from the center (5, 5) was measured. The average distance for the HAHV videos was 2.52 and the standard deviation was 0.48. The average distances for the HALV, LALV, and LAHV videos were 3.56 (standard deviation: 0.51), 2.48 (standard deviation: 0.37), and 2.56 (standard deviation: 0.42).

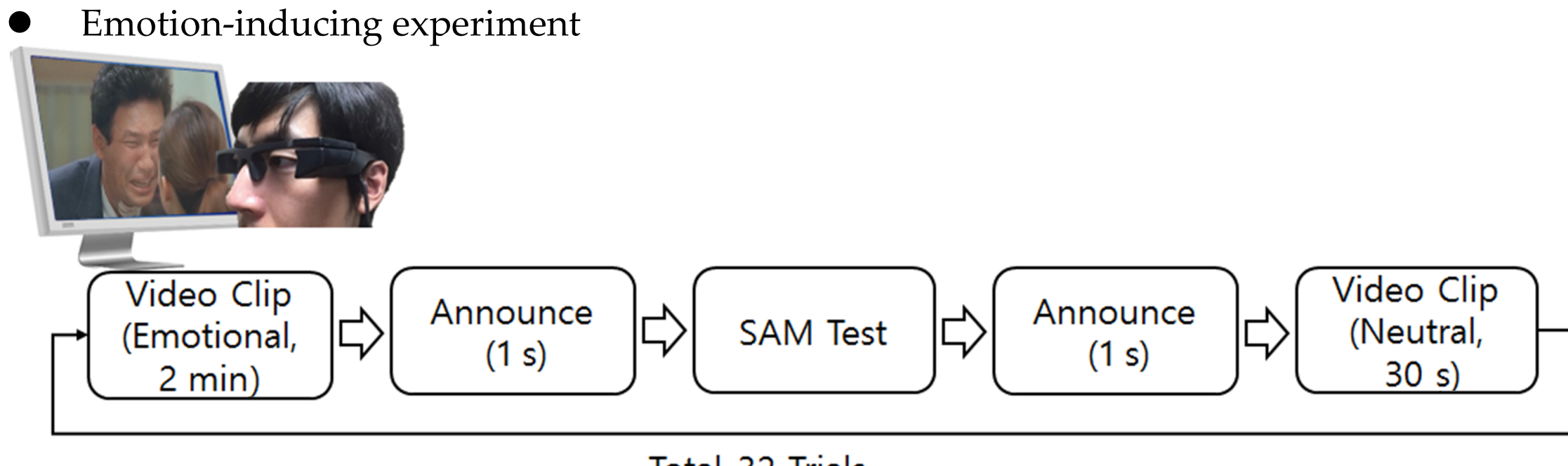

**Figure 8.** Flowchart of emotion-inducing experiment.

The emotion-inducing experiments were conducted using the selected stimuli. The experiment was organized into 32 trials. In each trial, the stimulus clips were shown for 2 min and the neutral clips were shown for 30 s. The neutral clips were shown to neutralize the emotional state between the trials. The emotions induced during the trials were counter-balanced to avoid label bias. E-Prime 2.0 [23] was used to present the stimulus.

Experiments were conducted using the stimulus clips. A total of 20 subjects participated in the experiment. A 1.7 m × 1.9 m × 3.0 m shield room was used for the experiment. The room contained a 19-inch monitor and a 2-way speaker. The distance between the subject's head and monitor was 1 m. To acquire data, the wearable device was worn by the subject, and the physiological signals and facial expressions were recorded as the subject was watching the clip. The physiological response data were recorded at 200 Hz and transferred via Bluetooth. The facial expressions were captured at 5 Hz and transferred via Wi-Fi. The data-acquisition procedure was manually programmed on MATLAB 2016b.

The subjects were in the age group of 21 to 35 years old, the average age was 26.7. The experimental protocol was carefully explained to the subjects upon arrival. The experiment consisted of 32 trials. In each trial, a 30-s-long neutral video clip was shown to neutralize the former emotional state of the subject. Then, 2-min-long emotional video clips were shown followed by a 2-s-long black screen. The subjects were requested to rest for 5 min after completing 16 trials to avoid the effect of stress on the physiological signal. The total trial was 1 h and 32 min long, but the entire process including preparation took ~2 h. All the procedures in the experiment were certified by the Institutional Review Board of KIST-2016.

*2.3. Data Processing*

- Physiological responses

The acquired raw data were processed in a traditional feature-based machine-learning manner. The emotion-related features were extracted from each raw data channel. The features were extracted from 120-s long physiological response data per trial. The observation length for feature extraction was 100-s long, and the features were extracted by moving the observed region for 6 s. Therefore, three feature vectors were acquired from each trial, and 96 feature vectors were acquired from a subject. Two kinds of features were extracted: statistical features and domain-related features.

The statistical features were extracted from the raw signal regardless of the acquired channel. The commonly used statistical features extracted from raw signal are: (1) average value, (2) standard deviation, (3) mean of absolute value of 1st difference, (4) mean of absolute value of 2nd difference, (5) ratio of mean of absolute value of the 1st difference and standard deviation, and (6) ratio of mean of absolute value of 2nd difference and standard deviation.

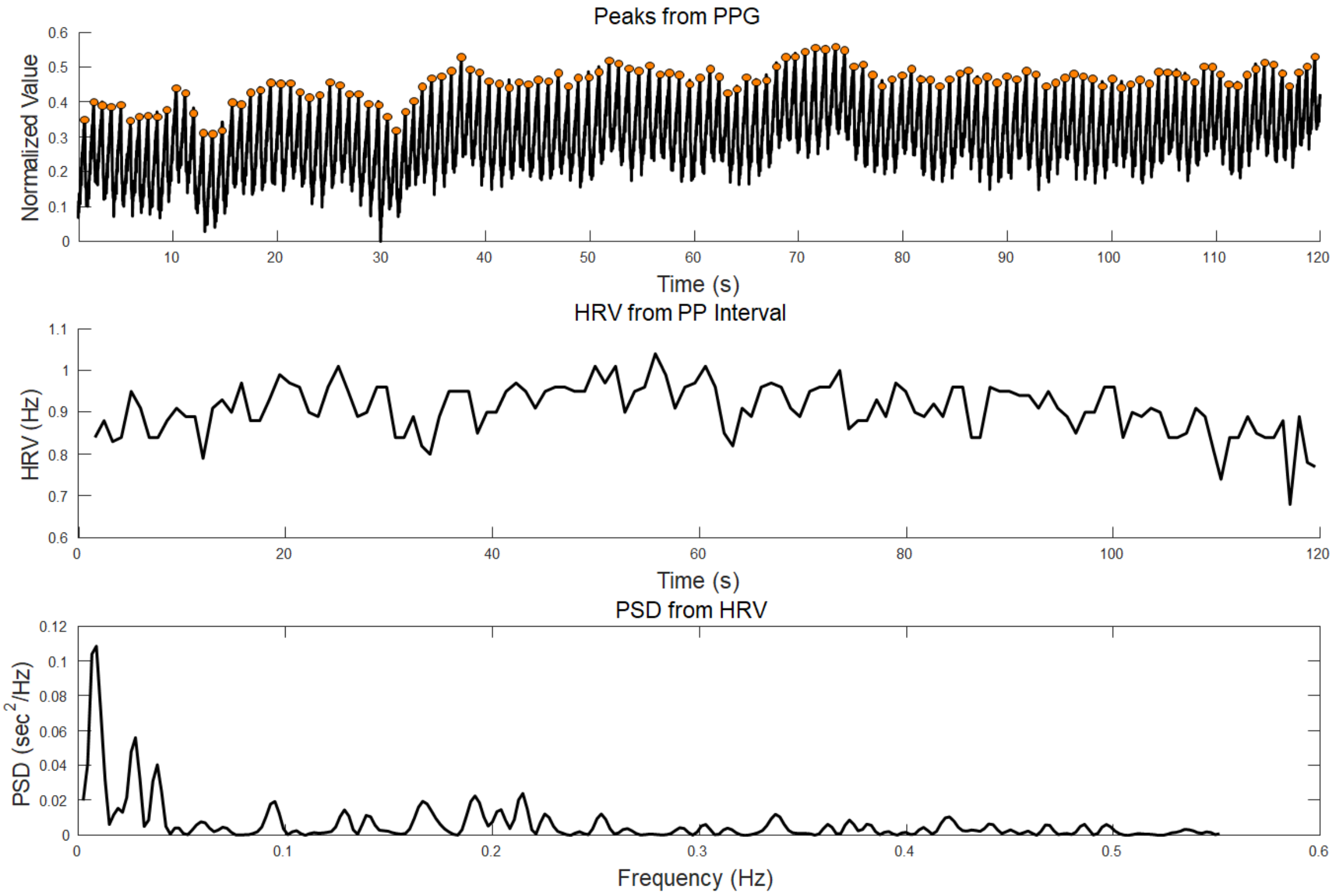

**Figure 9.** Extraction of HRV-related features from PPG signal.

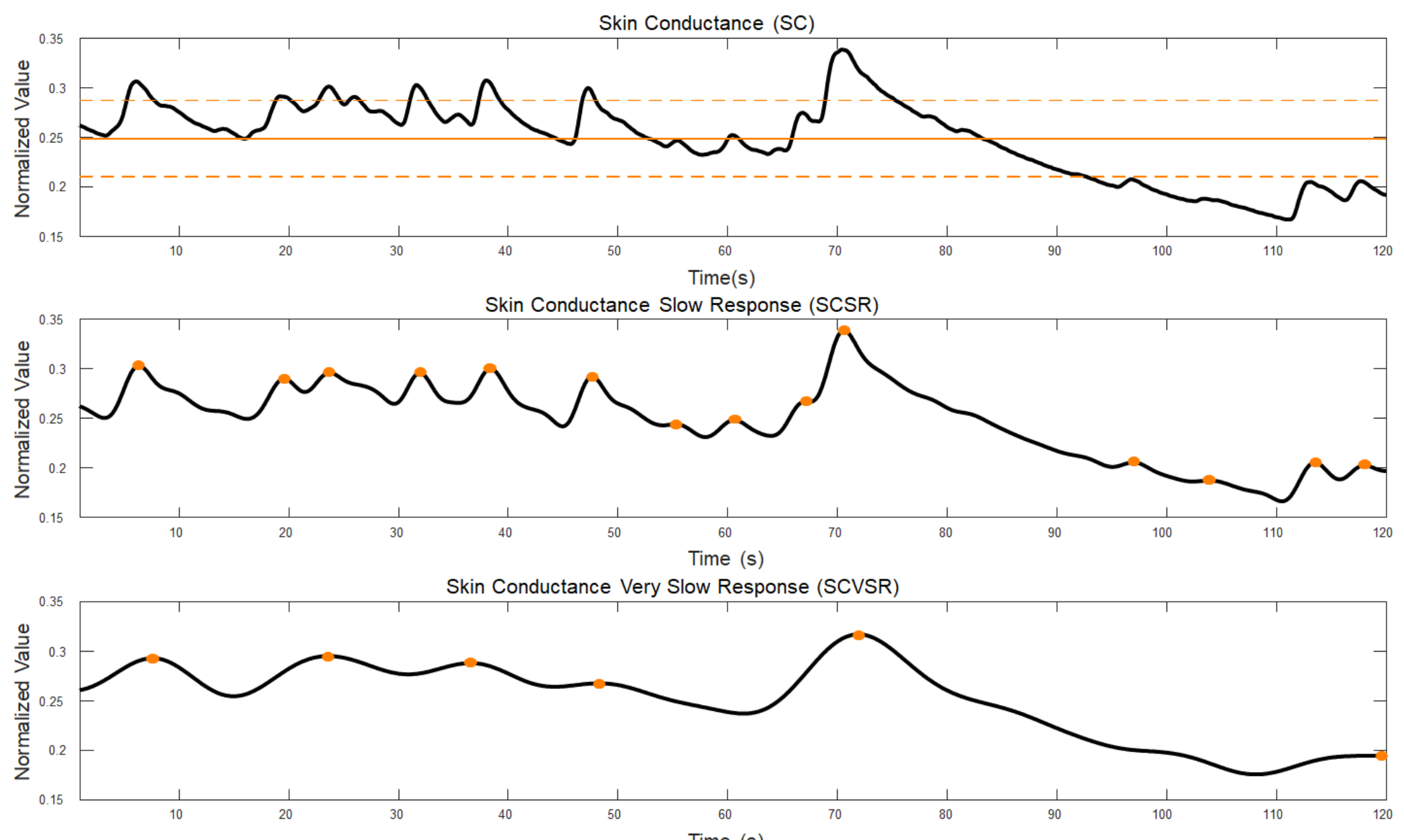

**Figure 10.** Extraction of SCR-related features from EDA signal.

**Table 1.** Types of domain-related features

| Channels | Features |
| --- | --- |
| PPG | (1) Average acceleration of pulse, (2) NN50, (3) RMSSD, (4) SDNN, (5) SDSD, (6) VLF, (7) LF, (8) HF |
| EDA | (1) Ratio of recovery time to signal length in SCSR, (2) Number of SCRs in SCSR, (3) Number of SCRs in SCVSR, (4) Average amplitude of SCRs in SCSR, (5) Average amplitude of SCRs in SCVSR |

The domain-related features were processed differently for each channel. For the PPG signal, the average acceleration [11] of the pulse and heart rate variability (HRV)-related features were extracted. The HRV features are extracted mainly from the peak-to-peak (PP) interval and power spectral density (PSD) of the PP interval. Figure 8 shows the PPG signal with PP interval and the corresponding PSD. The extracted features are presented in Table 1. In the EDA signal, most features were related to the skin conductance response (SCR). SCRs usually occur in 1 to 3 s intervals [24]. However, some SCRs occurred over longer intervals. Two low-pass filters were used to acquire the SCRs in different time resolutions. First, the raw signal was passed through a low-pass filter with 0.2-Hz cut-off frequency, which was called the skin conductance slow response (SCSR). Second, a low-pass filter with a 0.08-Hz cut-off frequency was applied, and the signal was called the skin conductance very slow response (SCVSR). The SCR-related features were extracted from each pre-processed signal.

After extracting the physiological response features, a feature-selection method was applied to filter the features that did not vary between emotional states. The selection algorithm used was ReliefF. [25]

- Facial expression

Data-processing based on facial expressions was conducted according to the Fisherface method [26]. We used the Fisherface method because it does not need facial landmarks. Other methods based on facial landmarks were hard to apply, because our facial expression data were captured from only a part of the face. The labels were determined according to the clip the subject watched, regardless of whether the subject actually expressed emotion. First, principal components analysis (PCA) was applied to reduce the dimensionality of the captured images. The principal components and the eigenvalues were extracted from the training dataset and sorted based on the magnitude of the eigenvalues. The eigenvectors were extracted in descending order until the sum of eigenvalues exceeded 90% of the total summation of eigenvalues. The pixel values of the captured images were mapped on to the selected components. The projected values were analyzed using linear discriminant analysis (LDA) to extract more discriminant features based on the user's emotional state. A weight matrix W was obtained from the LDA, which satisfies the following expression [26]:

$$\mathrm{W} = \arg\max_{W} \frac{|W^T W_{PCA}^T S_B W_{PCA} W|}{|W^T W_{PCA}^T S_W W_{PCA} W|}$$

$S_B$ is the scatter matrix between the average of data from each emotional class, and $S_B$ is the summation of scatter matrices extracted from each emotional class. $W_{PCA}$ is the matrix whose columns are the eigenvectors collected from the previous step. Therefore, two types of discriminant matrices — the weight matrix extracted from PCA and the one extracted from the LDA were extracted in the training phase to acquire discriminant features for emotion recognition.

- Data fusion and classification

The proposed device can only detect one emotional state; therefore, information from different channels needs to be combined for more accurate results. Information was combined in two ways: feature-level fusion and decision-level fusion.

In decision-level fusion, six classifiers were created with different parameters. Three classifiers from each channel were trained during a training session. The classifier models include quadratic discriminant analysis (QDA), the Gaussian mixture model (GMM), and k-nearest neighbors (KNN). In the test session, the emotional states were detected for all the classifiers regardless of the channels, and those with the maximum count was selected. For those with the same count, priority was given to the classifiers based on facial expression because they are empirically shown to have a higher accuracy than physiological responses. The number of mixtures of the GMM model was two, and the full covariance was used to optimize the parameters. The initial prior-probability of each emotional state was set to 1/(number of emotional states).

In feature-level fusion, feature vectors acquired from each channel were concatenated. PCA was used for the concatenated dataset to extract meaningful combinations of all the features. The acquired principal components were used in the test session.

It is important to match the features with the extraction time, because the observation lengths for each channel differ. For example, the camera module can capture a user's expression every 0.2 s, and a single captured image is enough to extract the features. However, regarding physiological responses, one needs to wait for approximately 90 s to collect enough signals to extract the feature. Therefore, the facial expression features should be matched with respect to the physiological response features. We matched average of facial expression features, which were acquired during the physiological responses acquisition process. When sufficient physiological responses were acquired, features from the average expression were extracted and matched with the physiological features.

## 3. Results

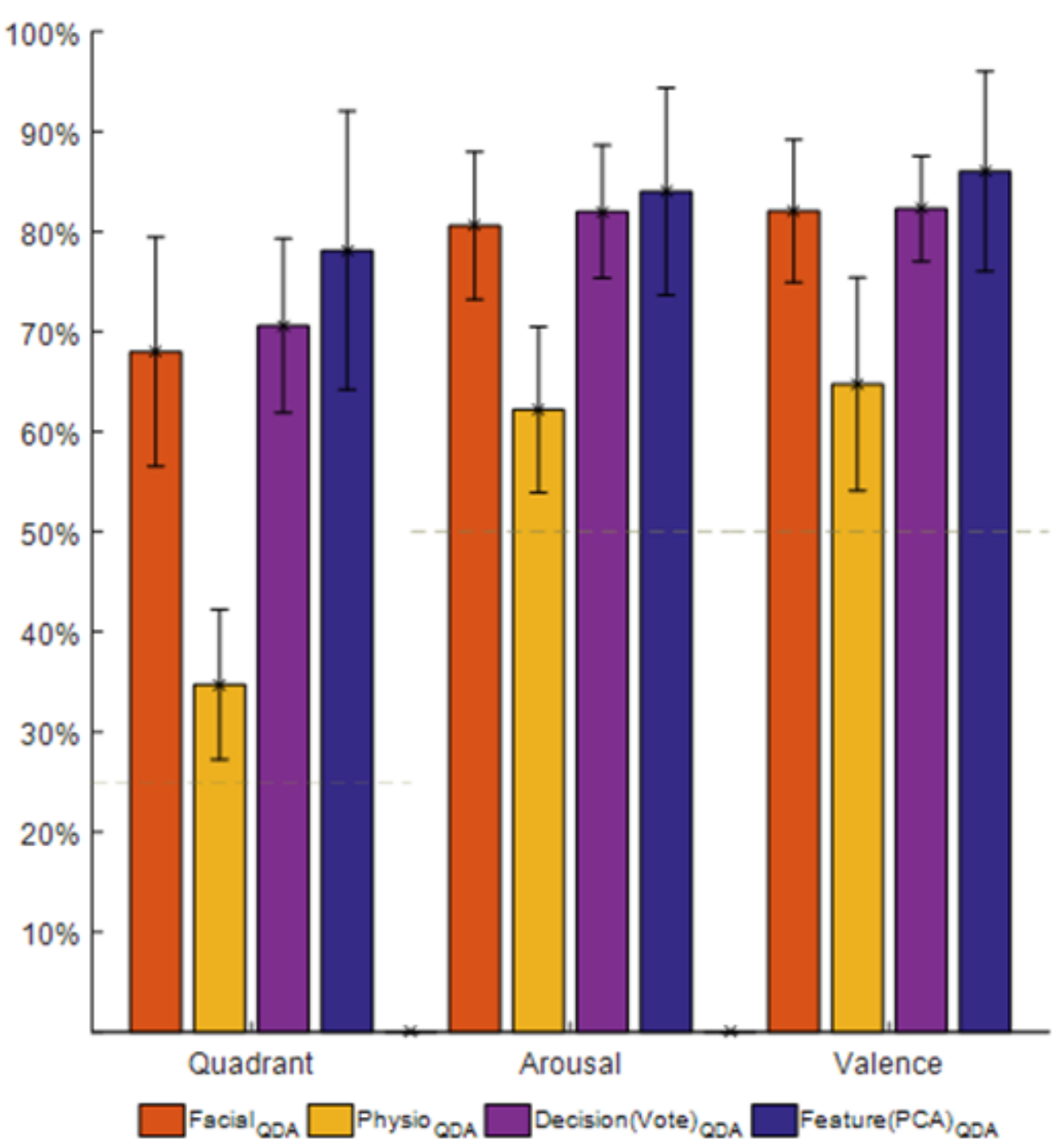

**Figure 11.** Classification results. The dashed line shows the chance level of each classification task.

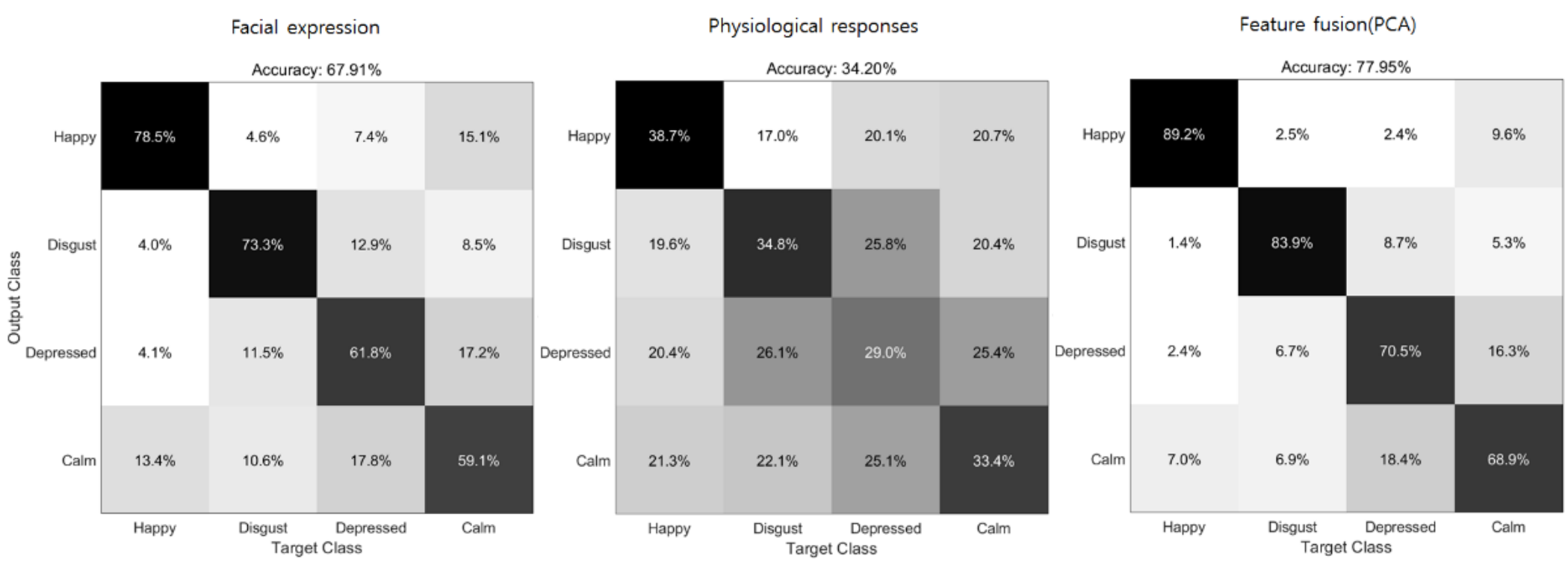

**Figure 12.** Confusion matrix for detecting quadrant emotion state.

The accuracy of classification using facial expressions was greater than that using physiological responses. The accuracy when using facial expression was 67.9% for estimating the quadrant emotional states, 80.6% for the arousal states, and 82.1% for the valence states. The accuracy when using physiological responses was 34.2% for estimating the quadrant states, 62.2% for the arousal states, and 64.8% for the valence states.

The recognition performance improved when the fusion methods were applied. The accuracy of the voting method and feature-fusion method improved by 2.7% and 10.1%, respectively, when compared with methods that used only one channel.

We found that the performance of the result of emotion recognition was considerably higher than the chance level. On the other hand, the physiological responses on the face show lower performance than the local facial expression. Nevertheless, when the local facial expressions and the physiological responses were combined, improvement in the performance of emotion recognition was observed.

## 4. Discussion

Existing wearable devices for emotion recognition that only use biological signals or facial expressions may yield poor results depending on the situation. To solve this problem, this study proposed a wearable device that uses both facial expressions and bio-signals. The proposed device exhibited good performance in an emotion recognition experiment. The results showed that the classification performance increased by 10% when the two modalities were combined, compared to the method using facial expressions alone. This performance improvement is consistent with the second research hypothesis presented in the introduction. The video clip stimuli used in the experiment are complex stimuli comprising image [27] and sound stimuli [28]. The increased classification performance for the combined channel compared to that for a single channel in the complex stimulus is consistent with the second hypothesis of the study that the multimodal method would be more useful in a complex situation.

This study extends the results of existing multimodal emotion recognition studies. In the existing study, biometric signals and facial expressions were used to perform emotion recognition at a standard measurement location. For example, the sweat response uses the EDA response of the finger and the facial expression uses a frontal view of the full face. Our study, on the other hand, measured user response in areas confined to the face, such as facial expressions around the eyes or sweat responses of the nasal skin. This location has not been measured in existing multimodal emotion recognition studies. Therefore, the modalities differ from those in existing multimodal emotion recognition studies. Our research provides support to multimodal emotion recognition researchers or facial wearable device developers in order to explore more real-world applicable modalities.

In terms of experimental methods, our study differs from previous wearable studies. Most of the experiments in existing facial wearable studies involved users intentionally making facial expressions. On the other hand, our study examined the naturally induced emotions of the user through the wearable device while the user is watching video. Therefore, our study used an experiment protocol that was closer to real-life scenarios than that in existing facial wearable studies.

The limitation of this study is that there was no scale to assess whether emotions were actually induced. We labeled the data according to the clip that the subject was watching while we were obtaining the data. However, owing to the nature of emotional response, there was no clear way to verify that the data was labeled accurately, which implies that there was no clear way to verify whether the user actually felt the emotion induced by the clip in the experiment. This problem is important in emotion recognition because incorrectly labeled data can affect the decision boundary of the classifier. Based on our experience, we recommended that the process of evaluating the dominance score in each trial within the experiment protocol should be included in further studies; additionally, we strongly recommend using the dominance scale to exclude unreliable data. Alternatively, the dominance score can be used as a soft target label for the data. For example, if the user felt an emotion in a certain trial with low dominance, the acquired data can be excluded or assigned a tiny weight when training the emotion recognition model.

We did not use a deep learning architecture because of the small number of trials. Unlike other physiological responses such as EEG signals, EDA and PPG signals need to be observed for long periods to obtain useful information. For example, the SCR features extracted from the EDA or the HRV features extracted from the PPG need to be observed for more than 1 min [29]. Therefore, it is difficult to extract many feature vectors from a single trial. Additionally, signals obtained from the wearable sensors were noisier than those from the laboratory sensors [30], which implies that more data is required in actual scenarios than in a laboratory setting. Hence, we did not use the deep learning model in this study.

## 5. Conclusion

In this study, to increase the reliability of emotion recognition, we propose a glasses-type wearable device that measures local facial expressions in addition to physiological responses. The facial expressions were acquired in an unobtrusive manner by using a camera, and the location of the biosensors in the wearable device was determined via signal measurement experiments. Experiments using video clips were conducted to measure the performance of the device. Our results show that the glasses-type wearable device can be used to estimate a user's emotional state accurately.

**Author Contributions:** conceptualization, Laehyun Kim, Jangho Kwon; methodology, Jangho Kwon; software, Jangho Kwon; validation, Jangho Kwon; formal analysis, Jangho Kwon; investigation, Jangho Kwon, Jihyeon Ha, Da-Hye Kim; resources, Jangho Kwon.; data curation, Jangho Kwon; writing—original draft preparation, Jangho Kwon, Laehyun Kim; writing—review and editing, Jangho Kwon, Laehyun Kim, Junwon Choi; visualization, Jangho Kwon; supervision, Leahyun Kim, Junwon Choi; project administration, Leahyun Kim; funding acquisition, Leahyun Kim"

**Funding:** This research was supported by Institute for Information & communications Technology Promotion(IITP) funded by the Korea government(MSIT) (No.2017-0-00432, Development of non-invasive integrated BCI SW platform to control home appliances and external devices by user's thought via AR/VR interface)

**Acknowledgments:** No acknowledgments.

**Conflicts of Interest:** The authors declare no conflict of interest.

## References

[1] J. Marín-Morales *et al.*, "Affective computing in virtual reality: emotion recognition from brain and heartbeat dynamics using wearable sensors," *Sci. Rep.*, vol. 8, no. 1, 2018.

[2] E. Reynolds, "Nevermind," *ACM SIGGRAPH 2013 Posters - SIGGRAPH '13*, p. 1, 2013.

[3] S. L. Stokes, "Emotional intelligence," *Inf. Syst. Manag.*, 2004.

[4] K. Wac and C. Tsiourti, "Ambulatory assessment of affect: Survey of sensor systems for monitoring of autonomic nervous systems activation in emotion," *IEEE Trans. Affect. Comput.*, vol. 5, no. 3, pp. 251–272, 2014.

[5] N. Nourbakhsh, Y. Wang, F. Chen, and R. A. Calvo, "Using galvanic skin response for cognitive load measurement in arithmetic and reading tasks," in *Proceedings of the 24th Australian Computer-Human Interaction Conference on - OzCHI '12*, 2012, pp. 420–423.

[6] C. Setz, B. Arnrich, J. Schumm, R. La Marca, G. Tröster, and U. Ehlert, "Discriminating stress from cognitive load using a wearable eda device," *IEEE Trans. Inf. Technol. Biomed.*, vol. 14, no. 2, pp. 410–417, 2010.

[7] R. A. Calvo and S. D'Mello, "Affect Detection: An Interdisciplinary Review of Models, Methods, and Their Applications," *IEEE Trans. Affect. Comput.*, vol. 1, no. 1, pp. 18–37, 2010.

[8] J. Scheirer, R. Fernandez, and R. W. Picard, "Expression Glasses®: A wearable device for facial expression recognition," *Media*, no. May, pp. 262–263, 1999.

[9] A. Gruebler and K. Suzuki, "Design of a wearable device for reading positive expressions from facial EMG signals," *IEEE Trans. Affect. Comput.*, vol. 5, no. 3, pp. 227–237, 2014.

[10] A. Teeters, R. El Kaliouby, and R. Picard, "Self-Cam: feedback from what would be your social partner," *ACM SIGGRAPH 2006 Res. …*, p. 2005, 2006.

[11] R. W. Picard, E. Vyzas, and J. Healey, "Toward machine emotional intelligence: Analysis of affective physiological state," *IEEE Trans. Pattern Anal. Mach. Intell.*, vol. 23, no. 10, pp. 1175–1191, 2001.

[12] K. Tehrani and M. Andrew, "Wearable Technology and Wearable Devices: Everything You Need to Know," *Wearable Devices Magazine*, 2014. [Online]. Available: http://www.wearabledevices.com/what-is-a-wearable-device/.

[13] P. Washington *et al.*, "A Wearable Social Interaction Aid for Children with Autism," in *Proceedings of the 2016 CHI Conference Extended Abstracts on Human Factors in Computing Systems - CHI EA '16*, 2016.

[14] T. Starner, "Project glass: An extension of the self," *IEEE Pervasive Comput.*, vol. 12, no. 2, pp. 14–16, 2013.

[15] A. Boxall, "Smith Lowdown Focus: Our First Take."

[16] K. Shin, Y. Kim, S. Bae, K. Park, and S. Kim, "A Novel Headset with a Transmissive PPG Sensor for Heart Rate Measurement," in *IFMBE Proceedings*, 2009, vol. 23, pp. 519–522.

[17] M. Z. Poh, N. C. Swenson, and R. W. Picard, "Motion-tolerant magnetic earring sensor and wireless earpiece for wearable photoplethysmography," *IEEE Trans. Inf. Technol. Biomed.*, vol. 14, no. 3, pp. 786–794, 2010.

[18] P. Celka, C. Verjus, R. Vetter, P. Renevey, and V. Neuman, "Motion Resistant Earphone Located Infrared based Heart Rate Measurement Device," in *2nd Int. Biomed. Eng.*, 2004, pp. 582–585.

[19] J. Kwon, D.-H. Kim, W. Park, and L. Kim, "A Wearable Device for Emotional Recognition Using Facial Expression and Physiological Response," *Eng. Med. Biol. Soc.*, pp. 5765–5768, 2016.

[20] M. van Dooren, J. J. G. G. J. de Vries, and J. H. Janssen, "Emotional sweating across the body: Comparing 16 different skin conductance measurement locations," *Physiol. Behav.*, vol. 106, no. 2, pp. 298–304, 2012.

[21] J. A. Russell, "A circumplex model of affect," *J. Pers. Soc. Psychol.*, vol. 39, no. 6, pp. 1161–1178, 1980.

[22] S. Koelstra *et al.*, "DEAP: A database for emotion analysis; Using physiological signals," *IEEE Trans. Affect. Comput.*, vol. 3, no. 1, pp. 18–31, 2012.

[23] I. Psychology Software Tools, "E-Prime 2.0." 2012.

[24] J. Cacioppo, L. G. Tassinary, and G. G. Berntson, *The Handbook of Psychophysiology*, vol. 44. 2007.

[25] M. Robnik-Šikonja and I. Kononenko, "An adaptation of Relief for attribute estimation in regression," in *ICML '97 Proceedings of the Fourteenth International Conference on Machine Learning*, 1997.

[26] P. N. Belhumeur, J. P. Hespanha, and D. J. Kriegman, "Eigenfaces vs. fisherfaces: Recognition using class specific linear projection," *IEEE Trans. Pattern Anal. Mach. Intell.*, vol. 19, no. 7, pp. 711–720, 1997.

[27] B. N. Lang, P.J., Bradley, M.M., & Cuthbert, "International affective picture system (IAPS): Affective ratings of pictures and instruction manual.," *Tech. Rep. A-8.*, 2008.

[28] J. Kim and E. André, "Emotion recognition based on physiological changes in music listening," *IEEE Trans. Pattern Anal. Mach. Intell.*, vol. 30, no. 12, pp. 2067–2083, 2008.

[29] F. Shaffer and J. P. Ginsberg, "An Overview of Heart Rate Variability Metrics and Norms," *Front. Public Heal.*, vol. 5, 2017.

[30] M. Ragot, N. Martin, S. Em, N. Pallamin, and J. M. Diverrez, "Emotion recognition using physiological signals: Laboratory vs. wearable sensors," in *Advances in Intelligent Systems and Computing*, 2018.